\documentclass[prl,twocolumn,showpacs,preprintnumbers,amsmath,amssymb,superscriptaddress]{revtex4}

\newtheorem{proposition}{Proposition}
\newtheorem{theorem}{Theorem}
\newtheorem{corollary}{Corollary}

\begin{document}

\title{Remote Preparation and Distribution of Bipartite Entangled States}  
\author{Gilad Gour}\email{ggour@math.ucsd.edu}
\affiliation{Theoretical Physics Institute, University of Alberta, Edmonton, Alberta T6G 2J1, Canada}
\affiliation{Department of Mathematics, University of California/San Diego, 
        La Jolla, California 92093-0112}
\author{Barry C. Sanders}
\affiliation{Institute for Quantum Information Science, University of Calgary, Alberta T2N 1N4, Canada}

\date{\today}

\begin{abstract} 
We prove a powerful theorem for tripartite remote entanglement 
distribution protocols 
that establishes an upper bound on the amount of
entanglement of formation that can be created between two single-qubit
nodes of a
quantum network.
Our theorem also provides an operational interpretation of concurrence
as a type of entanglement 
capacity.
\end{abstract}


\maketitle

Shared bipartite entanglement is a crucial shared resource for many quantum information tasks
such as teleportation~\cite{Ben93}, entanglement
swapping~\cite{Zuk93,Ben93,Bos99,Har00}, 
and remote state preparation (RSP)
\cite{Ben01,Ye04} that are employed
in quantum information protocols. In general different parties, or nodes of a quantum network (QNet)
share an entanglement resource, such as ebits (maximally entangled pure bipartite states),
which are consumed during the task. 
In practice, generating entangled states is expensive, but here we establish a protocol by which
a QNet requires only a single supplier of entanglement to all nodes who, by judicious
measurements and classical communication, provides the nodes with a unique
pairwise entangled state independent of the measurement outcome. Furthermore, we extend this
result to a chain of suppliers and nodes, which enables an operational interpretation
of concurrence~\cite{Woo98}.

In the special case that the supplier (whom we call ``Sapna'') shares bipartite states with 
two nodes (labeled ``Alice'' and ``Bob''), and such states are pure and maximally
entangled, our protocol corresponds to entanglement swapping. However, 
in the practical case that initial shared entanglement between
suppliers and nodes involves partially entangled or mixed states, we show that general 
local operations and classical communication (LOCC) by \emph{all} parties (suppliers and nodes)  
yields \emph{distributions} of entangled states between nodes. In general a distribution of
bipartite entangled states between any two nodes will include states that do
not have the \emph{same} entanglement (i.e.\ not all the states 
equivalent under LOCC between the nodes); thus we
name this general process remote entanglement 
distribution (RED). In our terminology entanglement swapping with partially entangled 
states~\cite{Har00} is a particular class of RED protocols.    
Here we identify which distributions of states (shared
between Alice and Bob) can or cannot be created by RED. In particular we prove a powerful
theorem that establishes, for the $(2\times 2)$-dimensional mixed case,
an upper bound on the entanglement of formation that can be produced
between Alice and Bob. We extend this result to the case of a linear chain
of parties that plays the role of suppliers and nodes; this extension provides
an operational interpretation of concurrence.

Then we discuss an especially interesting class of tripartite RED
protocols in which Alice and Bob (after LOCC by the three parties)
end up sharing a \emph{unique} bipartite entangled state, rather then a 
distribution of entangled states. In this scheme, Sapna not only
wishes to create entanglement between Alice and Bob, she wishes to
provide Alice and Bob with a single entangled state (which, in
general, is unknown to Alice and 
Bob~\cite{footnote:state}). When the initial bipartite
states (shared between Sapna and the two nodes) are partially entangled
$d\times d$ pure states, or belong to a particular non-trivial class of mixed
states, we provide a protocol for Sapna to remotely prepare a bipartite 
entangled state between Alice and Bob. In this protocol, Sapna performs a 
single orthogonal (von Neumann) measurement, then transmits $\log_2 d$ bits 
of classical information to 
Alice and $2\log_{2} d$ bits to Bob. Based solely on the classical
information received from Sapna, Alice and Bob perform local unitary 
operations to obtain the state that Sapna intends them to share.
Our protocol for remote preparation of bipartite entangled states (RPBES) works 
even when entanglement is insufficient for Sapna to simply teleport
qubits to Alice and Bob.

Our scheme for tripartite RED (including tripartite RPBES)
commences with a four-way shared state,
$\hat{\rho} _{1234}=\hat{\rho}_{12}\otimes\hat{\rho} _{34}$
for $\hat{\rho} _{12}$ and $\hat{\rho} _{34}$ bipartite entangled states,
and with Sapna holding shares 2 and 3, and Alice and Bob holding shares 1
and 4, respectively. Each share has a corresponding $d$-dimensional Hilbert space.
Alice, Bob and Sapna perform general LOCC (allowing classical communication amongst \emph{all} 
three parties) to create a set of outcomes 
\begin{equation}
\label{eq:outcome}
\mathcal{O}\equiv\{\hat{\sigma}_{14}^j=\text{Tr}_{23}\hat\sigma^j_{1234},
        Q_j;j=1,\ldots,s\}
\end{equation}
with $Q_j$ the probability that Alice and Bob
share the mixed state $\hat{\sigma}_{14}^j$, which is
obtained by reducing the four-way shared state $\hat\sigma^j_{1234}$ over Sapna's shares.

In the case of RPBES, $\hat{\sigma}_{14}^j$ represents the state obtained 
after a \emph{single} measurement performed by Sapna. Then, after Sapna broadcasts the measurement result
$j$, Alice and Bob each perform a single local unitary operation to
transform $\hat{\sigma}_{14}^j$ into a unique entangled state
(i.e. independent of $j$).
As mentioned above, this scheme for RPBES is always possible if $\hat{\rho} _{12}$ and $\hat{\rho} _{34}$ 
are partially entangled pure states or belong to a particular non-trivial class of mixed states
as we now show.

Let us begin by proving an important theorem that rules out certain
distributions (of bipartite states) from
being able to be created by general tripartite RED: this restriction is obtained via a bound
for the average concurrence of the resultant distribution shared by Alice and Bob in 
relation to the concurrences of the initial states Sapna has shared with each of Alice and Bob.
Concurrence for a pure bipartite state
$|\psi\rangle$ is $C\left(|\psi\rangle\right)\equiv\sqrt{2(1-{\rm Tr}\hat{\rho}_r^2)}$
\cite{Woo98,Run01,Min04} (with $\rho_r$ obtained by tracing the pure-state
density matrix $|\psi\rangle\langle\psi|$ over one of the two shares).
Concurrence for a mixed state $\hat{\rho}=\sum_i p_i |\psi_i \rangle\langle\psi_i |$ is defined as
the average concurrence of the pure states of the decomposition, minimized over
all decompositions of $\hat{\rho}$ (the convex roof): 
$C\left(\hat{\rho}\right)=\min\sum_i p_i C\left(|\psi_i \rangle\right)$.
(For an arbitrary state of two qubits the concurrence has been calculated 
explicitly~\cite{Woo98} and, recently, for higher dimensions a lower bound has been 
found~\cite{Min04}.)

\begin{theorem}
If Alice, Bob, and Sapna perform general LOCC on the initial four-qubit state 
$\hat{\rho}_{12}\otimes\hat{\rho}_{34}$
with outcome
$\{Q_{j}, \hat{\sigma}_{14}^{j}\}$, then
\begin{equation}
C_{14}\equiv\sum_{j=1}^{s}Q_{j}C(\hat{\sigma}_{14}^{j})\;\leq\;C_{12}C_{34},
\label{theorem1}
\end{equation}
with $C_{12}\equiv C(\hat{\rho} _{12})$ and
$C_{34}\equiv C(\hat{\rho} _{34})$.
\end{theorem}
\textbf{Proof:} Let us write $\hat{\rho} _{12}$ and $\hat{\rho} _{34}$ in their
\emph{optimal} decompositions
\begin{equation}
\hat{\rho}_{12} = \sum_{l=1}^4 p_{l}|\psi ^{(l)}\rangle_{12}\langle\psi ^{(l)}|
\;,\;\;\hat{\rho}_{34} = \sum_{l=1}^4 q_{l}|\chi ^{(l)}\rangle
_{34}\langle\chi ^{(l)}|;
\label{initial}
\end{equation}
we can always choose optimal
decompositions such that the four states
$|\psi^{(l)}\rangle _{12}$ have the same concurrence $C_{12}$ and all four states
$|\chi ^{(l)}\rangle _{34}$ have the same concurrence 
$C_{34}$~\cite{Woo98}. Thus, the Schmidt coefficients 
of the states $|\psi ^{(l)}\rangle _{12}$ and
$|\chi ^{(l)}\rangle_{34}$ do not depend on the index $l$:
\begin{align}
|\psi ^{(l)}\rangle _{12} & = \sqrt{\lambda _{0}}|0^{(l)}
0^{(l)}\rangle _{12}+\sqrt{\lambda _{1}}|1^{(l)}
1^{(l)}\rangle _{12}\nonumber\\
|\chi ^{(l)}\rangle _{34} & = \sqrt{\eta _{0}}|0^{(l)}
0^{(l)}\rangle _{34}+\sqrt{\eta _{1}}|1^{(l)}
1^{(l)}\rangle _{34}
\end{align}
with $\lambda _i$, $\eta _i$
the Schmidt coefficients of $|\psi ^{(l)}\rangle _{12}$, $|\chi ^{(l)}\rangle
_{34}$, respectively. The index $l$ in the states $\{|0^{(l)}\rangle
_{i}, |1^{(l)}\rangle _{i}\}$ represents four different bases for each
system $i=1,2,3,4$. Note that in this notation $C_{12}=2\sqrt{\lambda
  _{0}\lambda _{1}}$ and $C_{34}=2\sqrt{\eta _{0}\eta _{1}}$.   

Since the entanglement between Alice and Bob remains zero
unless Sapna perform a measurement, we assume that the first 
measurement is performed by Sapna and is described
by the Kraus operators $\hat{M}^{(j)}$ and their components
$M^{(j,ll')}_{mm',kk'}
\equiv {}_{23}\langle m^{(l)} m'^{(l')}|\hat{M}^{(j)}|k^{(l)}k'^{(l')}\rangle_{23}$,
with $k,k',m,m'=0,1$ and $l,l'=1,2,3,4$.
The density matrix shared between Alice and Bob after outcome $j$ occurs is
\begin{align}
\hat{\sigma}_{14}^{j} & = \frac{1}{Q_{j}}{\rm Tr}_{{}_{23}}\left(\hat{M}^{(j)\dag}\hat{\rho}_{12}\otimes
\hat{\rho}_{23} \hat{M}^{(j)}\right)\nonumber\\
& = \frac{1}{Q_j}\sum_{l,l'}\sum_{m,m'}p_lq_{l'}r^{(j,ll')}_{mm'}
|\phi^{(i,ll')}_{mm'}\rangle _{14}\langle \phi^{(i,ll')}_{mm'}|\;,
\label{deco}
\end{align} 
where $r^{(j,ll')}_{mm'} \equiv  \sum _{k,k'}\lambda _{k}\eta_{k'}|M^{(j,ll')}_{mm',kk'}|^{2}$,
\begin{equation}
|\phi^{(j,ll')}_{mm'}\rangle _{14} \equiv 
\frac{1}{\sqrt{r^{(j,ll')}_{mm'}}} 
\sum _{k,k'}\sqrt{\lambda _{k}\eta
  _{k'}}M^{(j,ll')}_{mm',kk'}|k^{(l)}
k'^{(l')}\rangle_{14}\;,\label{phi}
\end{equation}
and
$Q_{j}=\sum_{{}_{l,l',m,m'}}p_{l}q_{l'}r^{(j,ll')}_{mm'}$
is the probability to obtain an outcome $j$. Now, a 
direct calculation of $C\left(|\phi^{(j,ll')}_{mm'}\rangle
  _{14}\right)$ gives
\begin{align}
C&\left(|\phi^{(i,ll')}_{mm'}\rangle _{14}\right) =
\frac{2\sqrt{\lambda _{0}\lambda _{1}\eta _{0}
\eta _{1}}}{r^{(j,ll')}_{mm'}}\nonumber\\
&\times \left|M^{(j,ll')}_{mm',00}M^{(j,ll')}_{mm',11}
-M^{(j,ll')}_{mm',01}M^{(j,ll')}_{mm',10}\right|\;.
\end{align}
Therefore, since the concurrence of $\hat{\sigma}_{14}^{j}$ cannot exceed
the average concurrence of the decomposition in Eq.~(\ref{deco}), we have
\begin{align}
&  C_{14}\equiv \sum_{j=1}^{s}Q_{j}C\left(\hat{\sigma}_{14}^{j}\right)
\leq 2\sqrt{\lambda _{0}\lambda _{1}\eta _{0}
\eta _{1}}\sum _{l,l'}p_{l}q_{l'}\nonumber\\
& \times  \sum_{j}\sum_{m,m'}
\left|M^{(j,ll')}_{mm',00}M^{(j,ll')}_{mm',11}
-M^{(j,ll')}_{mm',01}M^{(j,ll')}_{mm',10}\right|\nonumber\\
& \leq\frac{1}{4} C_{12}C_{34}\sum_{j}{\rm Tr}\left(\hat{M}^{(j)\dag}\hat{M}^{(j)}\right)\;, 
\label{basic}
\end{align}
where the last inequality follows from the fact that 
$|ab-cd|\leq (|a|^{2}+|b|^{2}+|c|^{2}+|d|^{2})/2\;\forall\; a,b,c,d\in\mathbb{C}$.
Thus, from the completeness relation,
$\sum_{j}\hat{M}^{(j)\dag}\hat{M}^{(j)}=I$, we obtain Eq.~(\ref{theorem1}). 

Consider now the following LOCC: after Sapna's first measurement, she  
sends the result $j$ to Alice and Bob. Based on this result, Alice then performs a 
measurement represented by the Kraus operators $\hat{A}^{(k)}_{j}$ and sends
the result $k$ to Bob and Sapna. Based on the results $j,k$ from Sapna and Alice,
Bob performs a measurement represented by the Kraus operators $\hat{B}^{(n)}_{jk}$
and sends the result $n$ to Sapna. In the last step of this scheme, Sapna performs
a second measurement with Kraus operators denoted by $\hat{F}_{jkn}^{(j)}$ and sends 
the result $i$ to Alice and Bob. The final distribution of entangled
states shared between Alice and Bob is denoted by $\{N_{jkni},\hat{\sigma}
_{14}^{jkni}\}$, where $N_{jkni}$ is the probability for outcome
$j,k,n,i$.

Since the concurrence of any bipartite state, $|\psi\rangle_{14}$,
satisfies $C\left(\hat{A}_{j}^{(k)}\otimes\hat{B}_{jk}^{(n)}|\psi\rangle\right)
=|{\rm Det}(\hat{A}_{j}^{(k)})|\;
|{\rm Det}(\hat{B}_{jk}^{(n)})|\;C(|\psi\rangle)$, Eq.~(\ref{basic}) is now
replaced by
\begin{align}
&  C_{14}\equiv
\sum_{j,k,n,i}N_{jkni}C\left(\hat{\sigma}_{14}^{jkni}\right)
\leq\frac{1}{4}C_{12}C_{34}\sum_{j,k}\left|{\rm Det}(\hat{A}_{j}^{(k)})\right|\nonumber\\
& \times \sum_{n}
\left|{\rm Det}(\hat{B}_{jk}^{(n)})\right|\sum_{i}{\rm Tr}
\left(\hat{M}^{(j)\dag}\hat{F}^{(i)\dag}_{jkn}
\hat{F}^{(i)}_{jkn}\hat{M}^{(j)}\right)\;.
\label{nonbasic}
\end{align}
Moreover, from the geometric-arithmetic inequality we have
$\sum_{n}\left|{\rm Det}(\hat{B}_{jk}^{(n)})\right|\leq 
\frac{1}{2}\sum _{n}{\rm
  Tr}\left(\hat{B}_{jk}^{(n)\dag}\hat{B}_{jk}^{(n)}\right)=1$ and a similar
relation for $\hat{A}_{j}^{(k)}$. These results, together with
the completeness relation
$\sum_{i}\hat{F}^{(i)\dag}_{jkn}\hat{F}^{(i)}_{jkn}=1$,
lead us back to Eq.~(\ref{basic}).
As we can see, all operations that are performed by Alice, Bob and Sapna
after the first measurement by Sapna cannot increase the bound on $C_{14}\;\Box$.  

Theorem~1 concerns one supplier and two nodes, but in fact applies to one supplier
and \emph{any} pair of nodes; thus, the result of Theorem~1 is applicable to an arbitrarily
large QNet with one supplier and many nodes. In fact Theorem~1 can be extended to
more than one supplier, as stated in the following corollary.
\begin{corollary}
Consider an aligned chain of $N$ mixed bipartite two-qubit states, 
$\hat{\rho} _{01},\;\hat{\rho} _{12},...,\hat{\rho}_{N-1_{\;}N}$,
where the state 
$\hat{\rho}_{k-1_{\;}k}$
($k=1,2,...,N$) is shared between party $k-1$ and party $k$. If  
the $N+1$ parties perform LOCC on the initial state 
$\hat{\rho} _{01}\otimes\hat{\rho} _{12}\otimes\cdots\otimes\hat{\rho}_{N-1_{\;}N}$ with 
the resultant distribution of states between party $0$ and $N$ denoted by
$\{P_{j},\;\hat{\sigma}_{0N}^{j}\}$ ($P_{j}$ is the probability to have 
the state $\hat{\sigma}_{0N}^{j}$), then
\begin{equation}
C_{0N}\equiv\sum_{j}P_{j}C_{d}(\hat{\sigma}_{0N}^{j})\;\leq\;C_{01}C_{12}\cdots C_{N-1_{\;}N}\;,
\label{cor}
\end{equation}    
with $C_{k-1_{\;}k}\equiv C(\hat{\rho}_{k-1_{\;}k})$ ($k=1,2,...,N$). 
\end{corollary}

Theorem 1 and its corollary suggest an interpretation of the concurrence
as a form of \emph{entanglement capacity}. Until now 
concurrence has served as a powerful mathematical tool, but here 
we have introduced an operational description of the concurrence.
Furthermore, for two qubits, concurrence is equivalent to entanglement 
of formation; hence our theorem establishes an upper bound to the 
average amount of entanglement of formation that can be 
created by the supplier.

In the following, we will show that the equality in Eq.~(\ref{theorem1}) can always 
be achieved if both $\hat{\rho}_{12}$ and $\hat{\rho}_{34}$
are partially entangled and pure. Saturation of the bound is also possible if 
one of the states is maximally 
entangled and the other is any
mixed state (in which case the bound is saturated via
quantum teleportation~\cite{Ben93}).
Later we provide an example
showing that the bound saturates in some cases for
one state mixed and the other a \emph{partially} 
entangled pure state. It is not known, however, if saturation is always achievable.

\begin{proposition}
The equality in Eq.~(\ref{theorem1}) can always be achieved by RPBES 
if $\hat{\rho}_{12}$ and $\hat{\rho}_{34}$ are both $2\times 2$ bipartite pure states.
\end{proposition}
In order to prove that the equality in Eq.~(\ref{theorem1}) is achievable
for pure states, we establish a protocol for RPBES taking first 
$\hat{\rho}_{12}$ and $\hat{\rho}_{34}$ to be $d\times d$ pure states. Then 
we find that for $d=2$ our protocol saturates the inequality in 
Eq.~(\ref{theorem1}).

Working with partially entangled states is important because in the non-asymptotic regime
the process of concentration is expensive, and it is less expensive in terms of ebits consumed
(as well as classical bits~\cite{Gou04}) to work directly with partially entangled states~\cite{footnote:RSP}).
The protocol below enables Sapna to control the
amount of entanglement shared between Alice and Bob. In the
$(2\times 2)$-dimensional (pure) case the concurrence uniquely determines
the entanglement of the bipartite state. In this case
maximum concurrence corresponds to maximum possible entanglement.
However, for $d>2$ (or for mixed states), the concurrence of a 
$(d\times d)$-bipartite
(partially) entangled state is not sufficient to determine all the Schmidt 
coefficients. Thus, in this case, the optimal bipartite state that can
be prepared by Sapna is not unique. It depends on the choice taken for the
measure of entanglement; therefore, Sapna remotely prepares entangled
states according to the tasks Alice and Bob need to perform.

\emph{The RPBES Protocol.--}
Let the two pure densities be
expressed as $\hat{\rho}_{12}=|\psi\rangle_{12}\langle\psi|$ and 
$\hat{\rho}_{34}=|\chi\rangle_{34}\langle\chi|$ with $|\psi\rangle_{12}$
and $|\chi\rangle_{34}$ states in $d^2$-dimensional Hilbert spaces.
The initial states
$|\psi\rangle_{12}$ and $|\chi\rangle_{34}$ are expressed in the Schmidt
decomposition as
$|\psi\rangle_{12}  =\sum_{k=0}^{d-1}\sqrt{\lambda _{k}}|kk\rangle _{12}$ and
$|\chi\rangle_{34}  =\sum_{k=0}^{d-1}\sqrt{\eta _{k}}|kk\rangle _{34}$.
The steps of the protocol are as follows.
\textbf{(i)}~Sapna performs a projective measurement
\begin{equation}
\hat{P}^{(j,j')}=|P^{(j,j')}\rangle_{23}\langle P^{(j,j')}|,\,j,j'=0,1,...,d-1,
\label{eq:proj}
\end{equation}
with
\begin{equation}
|P^{(j,j')}\rangle_{23} \equiv   
 \frac{1}{d}\sum_{m,m'=0}^{d-1}\text{e}^{\text{i}\left[\frac{2\pi}{d^{2}}(dj+j')(dm+m')+\theta
 _{mm'}\right]}|mm'\rangle _{23}\;, 
\label{meas}
\end{equation}
with $\theta _{mm'}\in\mathbb{R}$ chosen freely. Note that the $d^2$
states $|P^{(j,j')}\rangle _{23}$ are orthonormal, regardless of the
choice of $\theta _{mm'}$.
\textbf{(ii)}~After the outcomes $j,j'$ have been obtained, the state of the system
can be written as $|P^{(j,j')}\rangle _{23}|\phi ^{(j,j')}\rangle _{14}$, where
\begin{align}
|\phi ^{(j,j')}\rangle _{14} & =  \sum _{m=0}^{d-1}\sum_{m'=0}^{d-1}
\sqrt{\lambda _{m}\eta _{m'}}\nonumber\\
& \times  \text{e}^{-\text{i}\left[\frac{2\pi}{d^{2}}(dj+j')(dm+m')+\theta _{mm'}\right]}|mm'\rangle _{14}\;.
\end{align}
\textbf{(iii)}~Sapna sends the results $j$ and $j'$ to Bob ($2\log_{2}d$ bits of
information) and 
the result $j'$ ($\log_{2}d$ bits of information) to Alice. 
Bob then performs the unitary operation
\begin{equation}
\hat{U}_{b}^{(j,j')}|m'\rangle
_{4}=\exp\left(\text{i}\frac{2\pi}{d^{2}}(dj+j')m'\right)|m'\rangle_4\;,
\label{eq:BobU}
\end{equation}
and Alice performs the unitary operation
\begin{equation}
\hat{U}_{a}^{(j')}|m\rangle_1=\exp\left(\text{i}\frac{2\pi}{d}j'm\right)|m\rangle _1\;.
\label{eq:AliceU}
\end{equation}
\textbf{(iv)}~The final state shared between Alice and Bob is
\cite{footnote:theta}
\begin{equation}
|F\rangle _{14}=\sum_{m=0}^{d-1}\sum_{m'=0}^{d-1}\exp\left
(-\text{i}\theta _{mm'}\right)\sqrt{\lambda_{m}\eta _{m'}}|mm'\rangle _{14}.
\label{final}
\end{equation}

For~$d=2$, Proposition 1 is proved by taking $\theta _{mm'}=\pi mm'$;
in this case the concurrence of $|F\rangle _{14}$ equals 
$4\sqrt{\lambda _{0}\lambda _{1}\eta _{0}\eta _{1}}=C_{12}C_{34}$, 
which is optimal (see Theorem~1). Moreover, if Alice and Bob know
the state prepared by Sapna, they can perform local unitaries to obtain
any state with the same concurrence. Thus,
\emph{any} $(2\times 2)$-dimensional bipartite pure state with concurrence
not greater than $C_{12}C_{34}$ can be prepared by Alice,
Bob, and Sapna performing LOCC. 

For $d>2$, if all $\lambda _{m}$ and $\eta _{m'}$ in Eq.~(\ref{final})
are equal to $1/d$, then the choice $\theta _{mm'}=2\pi mm'/d$ gives
a maximally entangled state. However, for different
values of $\lambda _{m}$ and $\eta _{m'}$, the optimal bipartite state
that can be prepared by Sapna depends on the choice of the 
entanglement measure.
For example, the concurrence of the $(d\times d)$ bipartite
state in Eq.~(\ref{final}) is
\begin{align}
C\left(|F\rangle _{14}\right) & =  
2{\Big\{}\sum_{k>k'}\sum_{m>m'}\lambda _{k}\lambda _{k'}\eta
  _{m}\eta _{m'}\nonumber\\
& \times 
\left|\text{e}^{\text{i}(\theta _{km}+\theta _{k'm'})}
-\text{e}^{\text{i}(\theta _{km'}+\theta _{k'm})}\right|^{2}{\Big\}}^{1/2}.
\label{eq:CF}
\end{align}
Unlike the $2\times 2$ case, for $d>2$ the term with the absolute
value in Eq.~(\ref{eq:CF}) cannot equal 2 for all 
$k,k',m,m'$. Thus, the values of $\theta _{km}$ that maximize 
$C\left(|F\rangle _{14}\right)$ depend explicitly on the Schmidt 
coefficients $\lambda_{k}$ and $\eta _{m}$. 

Our protocol can also be applied for mixed states. In general, for mixed states
$\hat{\rho}_{12}$ and $\hat{\rho}_{34}$, our protocol provides Alice and Bob 
with a \emph{distribution} of mixed states
rather then a unique state. As (for RPBES) Sapna wishes to produce a unique state $\hat{\sigma}_{14}$, 
we now establish a class of mixed bipartite states $\hat{\rho} _{12}$ and $\hat{\rho}_{34}$
for which our protocol yields a unique state.
We then give a specific example in the $(2\times 2)$-dimensional case, which we
show is optimal (maximum possible concurrence for the state shared by Alice and Bob).  

The two initial $(d\times d)$ bipartite density matrices can be expressed as
\begin{equation}
\hat{\rho}_{12}  =  \sum_{l=1}^{n}p_{l}|\psi ^{(l)}\rangle_{12}
\langle \psi ^{(l)}|\;\;,
\hat{\rho}_{34}  =  \sum_{l'=1}^{n'}q_{l'}|\chi ^{(l')}\rangle _{34}\langle \chi ^{(l')}|\;,
\end{equation} 
with $n,n'\leq d$ and 
\begin{equation}
|\psi ^{(l)}\rangle _{12}  =  \sum_{k=0}^{d-1}a_{k}^{(l)}|kk\rangle_{12}\;\;,
|\chi ^{(l')}\rangle _{34}  =  \sum_{k'=0}^{d-1}b_{k'}^{(l')}|k'k'\rangle_{34},
\end{equation}
with $a_k, b_{k'}\in\mathbb{C}$ and basis states
$|k\rangle _{i}$ independent of $l$ and $l'$; this 
characterizes the class of states containing $\hat{\rho}_{12}$ and $\hat{\rho}_{34}$.

Now, it can be shown that, 
after Sapna performs her measurement (\ref{eq:proj}),
and Bob and Alice perform the unitary operations of 
Eqs.~(\ref{eq:BobU}) and~(\ref{eq:AliceU}), the 
resultant shared state is
\begin{equation}
\hat{\sigma}_{14}=\sum_{l=1}^{n}\sum_{l'=1}^{n'}p_{l}q_{l'}
|\phi^{(ll')}\rangle _{14}\langle\phi ^{(ll')}|\;,
\label{fmx}
\end{equation}
with
\begin{equation}
|\phi ^{(ll')}\rangle _{14}=\sum_{k=0}^{d-1}\sum_{k'=0}^{d-1}
a_{k}^{(l)}b_{k'}^{(l')}\text{e}^{-\text{i}\theta _{kk'}}|kk'\rangle_{14}\;.
\label{mixr}
\end{equation}

We conclude with a simple interesting example. Suppose Alice shares with Bob the
$(2\times 2)$-dimensional pure state $|\psi\rangle _{12}=\sqrt{\lambda _0}|00\rangle_{12}
+\sqrt{\lambda _1}|11\rangle _{12}$ and Sapna shares with Bob the 
$(2\times 2)$-dimensional mixed state 
\begin{equation}
        \hat{\rho}_{34}=q|\chi^{(+)}\rangle_{34}\langle\chi ^{(+)}|
                +(1-q)|\chi^{(-)}\rangle_{34}\langle\chi^{(-)}|\;,
\end{equation}
where $0\leq q\leq 1$ and 
$|\chi^{(\pm)}\rangle_{34}=(1/\sqrt{2})(|00\rangle _{34}\pm |11\rangle _{34})$.
The concurrence of $|\psi\rangle _{12}$ is $2\sqrt{\lambda_{0}\lambda _{1}}$ and
the concurrence of $\hat{\rho}_{34}$ is equal to $|2q-1|$~\cite{Woo98}.
It is easy to see that both $|\psi\rangle _{12}$ and $\hat{\rho}_{34}$ belong to the 
class of density matrices described above. In this simple example, it is possible to 
calculate the concurrence of the final mixed state $\hat{\sigma}_{14}$ given in 
Eq.~(\ref{fmx}):
$C_{14}=|2q-1|\sqrt{\lambda _{0}\lambda _{1}}\left|\text{e}^{\text{i}(\theta _{00}+\theta _{11})}
-\text{e}^{\text{i}(\theta _{01}+\theta _{10})}\right|$.
Therefore, for $\theta _{kk'}=\pi kk'$, $k,k'=0,1$, we obtain 
$C_{14}=C_{12}C_{34}$: the bound in Theorem 1 is saturated. Thus, 
in this example the protocol is optimal and Bob can prepare the mixed bipartite 
state $\hat{\sigma} _{14}$ with \emph{any} value of concurrence between
0 and $C_{12}C_{34}$. 

In summary, we have introduced a protocol for a QNet that allows a single
supplier, who first shares entanglement with all nodes of the QNet (which may be
partially entangled pure states or a particular class of mixed states),
to provide any pair of nodes in the QNet with a single bipartite entangled state.
We have also proved a powerful theorem for tripartite RED protocols that establishes 
an upper bound on the amount of
entanglement of formation that can be created between two single-qubit nodes of the QNet.
We have also proven that it is possible (in some cases) to saturate the concurrence bound 
in the theorem if one state is pure (even if it is partially entangled), and the other is mixed. 
Our theorem also provides an operational interpretation of concurrence as a type of entanglement capacity.

\textbf{Acknowledgments:} GG acknowledges support by
the Killam Trust, the DARPA QuIST program under contract F49620-02-C-0010, 
and the National Science Foundation (NSF) under grant ECS-0202087. 
BCS acknowledges support from iCORE and CIAR.


\begin{references}

\bibitem{Ben93}
C.~H.~Bennett \emph{et al},
\prl \textbf{70}, 1895 (1993).

\bibitem{Zuk93}
M.~\.{Z}ukowski \emph{et al},
\prl \textbf{71}, 4287 (1993).

\bibitem{Bos99}
S. Bose \emph{et al},
\pra \textbf{57}, 822 (1998); 
\textbf{60}, 194 (1999);
B.-S. Shi \emph{et al},
\pra \textbf{62}, 054301 (2000).

\bibitem{Har00}
L.~Hardy and D.~D.~Song, \pra \textbf{62}, 052315 (2000).

\bibitem {Ben01}
C. H. Bennett \emph{et al},
\prl \textbf{87}, 077902 (2001).


\bibitem {Ye04}
M.-Y. Ye \emph{et al},
\pra \textbf{69}, 022310 (2004);
D. W. Berry, \pra (accepted).

\bibitem{Woo98}
W.~K.~Wootters, \prl \textbf{80}, 2245 (1998).

\bibitem {footnote:state} 
        However, if they do know the state, Sapna, Alice and Bob
        will be able to prepare a larger class of states.

\bibitem{Run01}
P.~Rungta \emph{et al},
\pra \textbf{64}, 042315 (2001).

\bibitem{Min04}
F.~Mintert \emph{et al},
\prl \textbf{92}, 167902 (2004).

\bibitem{Ben96}
C.~H.~Bennett \emph{et al},
\pra \textbf{54}, 3824 (1996). 

\bibitem{Gou04}
G.~Gour, \pra (accepted), arXiv:quant-ph/0402133.

\bibitem {footnote:RSP}
        RSP with partially entangled states has also been studied~\cite{Ye04};
        here, however, we are concerned with four-node sharing (Sapna with two shares 
        and Alice and Bob each with one share).

\bibitem{footnote:theta}
        The state~$|F\rangle_{14}$ is separable for $\theta _{mm'}=0$. 

\end{references}
\end{document}